\documentstyle[12pt]{article}
\newcommand{\sect}[1]{\setcounter{equation}{0}\section{#1}}

\begin{document}
%%\textwidth 150mm
%%
%\begin{flushright}
%{UT-KOMABA-95/3}\\
%{NSF-ITP-95-33}\\
%{\sl April, 1995}\\
%\end{flushright}
\vskip 2mm
\begin{center}
{\large\bf  COMPLEX RANDOM
MATRIX MODELS WITH POSSIBLE APPLICATIONS TO SPIN-IMPURITY
SCATTERING IN QUANTUM HALL FLUIDS
}
\end{center}
\vskip 3mm
\begin{center}
 S. Hikami$^{1}$ and A. Zee$^{2}$
\end{center}
\vskip 2mm
\begin{center}
%$^{1}$
%{Laboratoire de Physique Th\'eorique, Ecole Normale
%Sup\'erieure}\\
%{24 Rue Lhomond, 75231Paris, France}
%\vskip 2mm
{Department of Pure and Applied Sciences$^{1}$
}\\
{University of Tokyo, Meguro-ku, Komaba, Tokyo 153, Japan}\\
\vskip 2mm
Institute for Theoretical Physics{$^{2}$
}\\
{University of California
Santa Barbara, CA 93106, USA}\\
\end{center}
\vskip 5mm
\begin{abstract}

We study the one-point and two-point Green's functions in a
complex random matrix model to sub-leading orders in the
large $N$ limit. We  take this complex matrix model as a model
for the
two-state scattering problem, as applied to spin dependent
scattering of impurities in quantum Hall fluids. The density of
state shows a singularity at the band center due to reflection
symmetry. We also compute the one-point Green's function for a
generalized situation by putting random matrices
on a lattice of arbitrary dimensions.

\end{abstract}
\newpage

\sect{Introduction}

Recently,  matrix modes have been studied in
various contexts. In particular, in a series of papers [1-4], the
universality of the connected
two-point correlation function was discussed.
These studies focussed on $N$ by $N$ Hermitian matrices with
$N$
approaching infinity.

In this paper, we investigate the behavior of the Green's
functions of a complex matrix model. To leading order in
the large $N$ limit, complex matrix model and
Hermitian matrix model behave similarly. For instance, the
density of state,
for
Gaussian randomness, obeys Wigner's  semi-circle law in both
cases.
However, in subleading orders, complex random matrix
and Hermitian random matrix behave differently.  In the
language of
random
surfaces, a random complex matrix model represents, due to the
complex conjugate pairs, a surface made of plaquets with two
colors, like a red and black checkerboard, with the rule that a
given plaquette can only be  glued to a plaquette of a different
color.

Our motivations stem from possible applications of the complex
random matrix
model to physical problems involving scattering on impurities.
In particular, recently Hikami, Shirai, and Wegner [5] has
proposed a model for impurity scattering in quantum Hall fluids
in the spin degenerate case. For certain quantum Hall samples,
disorder
broadening can be much larger than the Zeeman splitting
between spin up
and spin down electrons, in which case the spin up and spin
down electrons
have the same energy. In [5] the further simplification is made
that when a
spin up electron scattering on an impurity becomes a spin down
electron,
and vice versa. This is known as the ``strong spin orbit" case, in
which case
it is known that
an extended state
appears at the band center of the lowest Landau level
with white noise
Gaussian random scattering and that the density
of state shows a singularity at the band center [5-9].
Although this problem has been simulated numerically [9],
the nature of the singularity of the density of state
remains unclear. Another example involving scattering between
two states occurs in high temperature superconductors, in which
the conducting plane contains  two different
sites, copper and oxygen. The
density of state also shows a singular behavior at the band center.

More generally, the problem of scattering between $C$ sectors (with $C=2$
in the example mentioned above) is of interest.
It turns out that a class of matrix models called ``lattices of
matrices" and
studied by
Br\'ezin
and Zee [3] is  relevant for this class of  problem.
 In
these models, random matrices are placed on a lattice of
arbitrary
dimension [10]. It was shown that in the large $N$ limit,
various
correlation
functions can be determined. The two-state scattering problem
considered
here corresponds to the simple case of a lattice consisting of two
points.
Following the analysis of [3] we can readily generalize the two-
state
scattering problem to an arbitray lattice.

In this paper, we show that  the
one-point Green's function is singular in the next-to-leading order
in
the large $N$ expansion.
We first evaluate the one-point Green's function
and the density of state to order $1/N^2$ by the
diagrammatic
method for one matrix model.
 The singular behavior of the one-point  Green's function
in
order $1/N^2$  has been
noticed in the literatures [11], but our discussion of this
singularity in
the
context of the two-state scattering problem in condensed matter
physics may be new.
Next we  discuss the origin of this spurious singularity
by the orthogonal polynomial method.

On the other hand, if we fix $N$ to be large but finite, and let the
energy
$E$ go to zero, the
density of state oscillates and eventually goes to zero. This
phenomenon is
due to energy-level repulsion.
The discussion here is somewhat reminiscent of the double
pole encountered in the
connected correlation function in the
Hermitian matrix model when evaluated in the diagrammatic
approach [2]. In the diagrammatic approach, to calculate the
Green's functions we select diagrams by letting $N$ go to
infinity first. We then obtain the correlation function by taking the
imaginary part of the two-point Green's function. The connected
correlation function has a double pole as its two arguments
approach each
other. In contrast, in the orthogonal polynomial approach, we in
effect take
the imaginary part of the two-point Green's function first, and then
let $N$
go to infinity. The connected two-point correlation calculated in
this way
does not have a double pole when its two arguments approach
each other.
However, when we smooth out the short distance oscillations of the
correlation function by averaging it appropriately, we recover the
double
pole obtained in the diagrammatic approach [1].

This paper is organized as follows. In section two, the two-state
scattering problem is formulated as a  complex
matrix model. We evaluate the one point Green's function
diagrammatically and
discuss the singularity of the density of state in $d=0$.
In section three, we study the complex matrix model further using
the orthogonal polynomial method.
In section four, we develop the analysis for
general dimensions, using
the ``lattice of matrices" formulation given in [3]. Using the
diagrammatic
expansion [12,13], we obtain the expression for the
one-point Green's function.
We compare our results
with that obtained in previous studies [14-16].

%**********************************************
%     section 2.
%**********************************************%
\sect{ Matrix model formulation of two-state scattering }

   In the standard Hermitian one-matrix model, the one-point
Green's
function $G(z)$  is defined by
%(2.1)
\begin{equation}
   G(z) = {1\over N} < {\rm Tr} {1\over{ z - \varphi }} >
\end{equation}
where the average is taken with the probability distribution
$P(\varphi)$

%(2.2)
\begin{equation}
  P(\varphi) = {\exp}[ - N{\rm Tr} V(\varphi) ]
\end{equation}
The $N\times N$ matrix $\varphi$ is  Hermitian. In the Gaussian
case, we
have
%(2.3)
\begin{equation}
    V(\varphi) = {1\over {2}} {\rm Tr}  \varphi^{2}
\end{equation}
More generally, we have, for
example, as
%(2.4)
\begin{equation}
 V(\varphi) = {1\over{2}} {\rm Tr}\varphi^{2} + {g\over{N}}{\rm
Tr}{\varphi^{4}}\end{equation}
For application to disordered systems, the random matrix
$\varphi$ is
interpreted as the Hamiltonian.

 We start with the simplest case of two-state scattering. The model
is
described by the random Hamiltonian
[3],
%{2.5}
\begin{equation}
      H = \left(\matrix{H_{1}&\varphi^{\dagger}\cr
                 \varphi&H_{2}\cr}\right)
\end{equation}
taken from the Gaussian distribution $P(H)$,
%(2.6)
\begin{equation}
     P(H) = {1\over{Z}} {e}^{-N{\rm Tr}[{1\over{2}}(m_{1}^2
H_{1}^{2}
+ m_{2}^2 H_{2}^{2}) + m^2 \varphi^{+}\varphi]}
\end{equation}
The matrices $H_1$ and $H_2$ are Hermitian while $\varphi$ is
complex.
In [3] this Hamiltonian was taken to describe a system with two
sectors (C=2).
Here we can think of the two sectors as representing as the spin
up and spin
down sectors in a spin-dependent quantum Hall system. We may
also think
of possible applications to the double-layered quantum Hall
system [17,18].

We now go to the model of (2.6) by letting
$m^2_{1}=m^2_{2}=\infty$, so that $H_1$ and $H_2$ are
suppressed. This model is for the off-diagonal disorder.
This same complex matrix model is considered also in [19].
(We will treat the more general case with $H_1$ and $H_2$ non-
zero later
in this paper.) We have
%(2.7)
\begin{equation}
     H = \left(\matrix{0&\varphi^{\dagger}\cr
               \varphi&0\cr}\right)
\end{equation}
Notice that there exists a matrix
%(gamma)
\begin{equation}
     \Gamma = \left(\matrix{I&0\cr
               0&-I\cr}\right)
\end{equation}
such that
%(anti)
\begin{equation}
     \{\Gamma, H\}= 0
\end{equation}
This implies that if $\psi$ is an eigenstate of $H$ with eigenvalue
$E$, then
$\Gamma \psi$ is an eigenstate with eigenvalue $-E$. Thus,
eigenvalues of
$H$ come in pairs. Due to level repulsion, around $E=0$, there
should be
a "hole" of width of order $1/N$ in the density of state $\rho(E)$. As
$N$
goes to infinity, this hole disappears and $\rho(E)$ should become
smooth.

To proceed, we calculate the   one point
Green's function, with $P = Z^{-1} exp(-{1\over 2} tr H^2)$ where
$H$ is given by (2.7).
%{2.8}
\begin{equation}
      G(z) = < {1\over {2N}}{\rm Tr} ({1\over{z - H}})
>
= < {1\over N}{\rm Tr} ({z\over{z^{2} - \varphi^{\dagger}\varphi}})
>
\end{equation}

To leading order in the large $N$ limit, we obtain easily
%{2.9}
\begin{equation}
        G(z) = {z - \sqrt{z^{2} - 4}\over{2}}
\end{equation}
using for example the diagrammatic approach of [2].
The density of state $\rho(E)=-{1\over{\pi}}{\rm Im }G(E)$ is given by
the semi-circle law
%(2.10)
\begin{equation}
        \rho(E) = {1\over{2\pi}}\sqrt{4 - E^{2}}
\end{equation}
We denote (2.11) by $G_{0}(z)$ hereafter.

To leading order, as expected, the density of state $\rho (E)$ is
smooth,
without any singularity at $E = 0$. However,
if we go to order $1/N^{2}$, we will find a divergent term in the
one point Green's function. Henceforth, we will work
with the Gaussian distribution. (We expect that the singularity at
$E=0$
will
occur also for a non-Gaussian distribution.)

  Using the  diagrammatic approach, we can readily evaluate the
Green's function to order  $1/N^{2}$ [2,12,13]. We decompose the
self-
energy $\Sigma
(z)$ into two parts $\Sigma_{a}$
and $\Sigma_{b}$,  obtained by breaking the solid line
(quark line in the terminology of [2]) in the diagrams $D_{a}$ and
$D_{b}$, respectively.

%\vskip 2mm
% (Fig.1: $\Sigma_{a1},\Sigma_{a2},\Sigma_{b1},\Sigma_{b2}$)
%\vskip 2mm
In the simplest two-state scattering case considered here, in which
scattering from
1 to 2 and from 2 to 1 occurs, but not from 1 to 1 or from 2 to 2,
we see
that
the numbers  $n$ and $m$ appearing in the diagram $D_{a}$,
describing
the
number of rungs in the gluon ladders, must be even. In the
diagram
$D_{b}$,
$n_{1},n_{2}
$ and $n_{3}$ must be all even, or all odd. We have
from  diagram $D_{a}$ two terms as follows.  Denoting the even
number of rungs in the ladder  by $2n$ and
$2m$, we have
%(2.11)
\begin{eqnarray}
   \Sigma_{a} &=& \sum_{n,m=1}^{\infty} G_{0}^{4n+4m +1}
    + \sum_{n,m=1}^{\infty} G_{0}^{4n + 4m +1} (2n
-1)\nonumber\\
    &=& {2G_{0}^{9}\over{( 1 - G_{0}^{4})^{3}}}
\end{eqnarray}
The factor of $(2n -1)$ is due
to the number of ways of inserting a cut inside the ladder.
Similarly, for  $\Sigma_{b1}$, we have
%(2.12)
\begin{eqnarray}
   \Sigma_{b1} &=& \sum_{n,m,l=1}^{\infty}
    G_{0}^{4n+4m+4l+1} + \sum_{n,m,l=1}^{\infty}
    G_{0}^{2(2n-1)+2(2m-1)+2(2l-1)+1}\nonumber\\
   &=& {G_{0}^{7}( 1 + G_{0}^{6})\over{( 1 - G_{0}^{4})^{3}}}
\end{eqnarray}
For $\Sigma_{b2}$, in which the cut appears inside the ladder, we
have
%(2.13)
\begin{eqnarray}
   \Sigma_{b2} &=& \sum_{n,m,l=1}^{\infty}
    G_{0}^{4n + 4m + 4l +1} (2n -1)\nonumber\\
   &+& \sum_{n,m,l=1}^{\infty}
    G_{0}^{2(2n-1)+2(2m-1)+2(2l-1)+1} ( 2n - 2)\nonumber\\
    &=& {2G_{0}^{7}( 1 + G_{0}^{6})\over{( 1 - G_{0}^{4})^{4}}}
     - { 2 G_{0}^{7} + G_{0}^{13} \over{( 1 - G_{0}^{4})^{3}}}
\end{eqnarray}
Thus we get
%(2.14)
\begin{eqnarray}
   \Sigma_{b} &=& \Sigma_{b1} + \Sigma_{b2}\nonumber\\
    &=& {2G_{0}^{7}( 1 + G_{0}^{6})\over{(1 - G_{0}^{4})^{4}}}
     - {G_{0}^{7}\over{(1 - G_{0}^{4})^{3}}}
\end{eqnarray}
Adding the two terms $\Sigma_{a}$ and $\Sigma_{b}$, we
have
%(2.15)
\begin{equation}
    \Sigma = {G_{0}^{7}\over{(1 + G_{0}^{2})^{2}( 1 -
G_{0}^{2})^{4}}}
\end{equation}
where $G_{0}$ is the one-point Green's function (2.9) evaluated to
leading
order
in the large $N$ limit.
After including an extra factor $1/(1 - G_{0}^{2})$ for the external
legs, we obtain the one-point Green's function to order $1/N^2$
%{2.16}
\begin{equation}
  G(z) = G_{0} + {1\over {N^{2}}}{G_{0}^{7}\over{(1 +
G_{0}^{2})^{2}
(1 - G_{0}^{2})^{5}}} + O({1\over{N^{4}}}).
\end{equation}

Since  $G_{0}(z)$ goes to $(-i)$ as $z\rightarrow 0$, the factor $1/(1
+
G_{0}^{2})$ diverges. Using $ 1 + G_{0}^{2} = z G_{0}$, we have
%{2.17}
\begin{equation}
  G(z) = G_{0}(z) + {1 \over{ N^{2} z^{2} ( z^{2} - 4 )^{5/2}}} + O({1\over
N^4})
\end{equation}
Thus, to order $1/N^{2}$, the Green's function $G(z)$ has a
singular
imaginary part $
i/(32 N^{2} z^{2})$ as $z\rightarrow 0$. This singularity is
related  to the reflection symmetry (parity symmetry) as
mentioned earlier.
The density of state diverges like
$\rho(E) \rightarrow 1/(32\pi N^{2}E^{2})$.  Apparently, this double
pole
is too singular,
since the integral of the density of
state should be one.

  We consider next the
connected
two-point correlation function $\rho_{2c}(z,w)$, which may be
obtained
from
the connected two-point Green's function \\
$G_{2c}(z,w)$ as shown in [1,2].
We have from a diagrammatic analysis [2-4],
%(2.18)
\begin{eqnarray}
 &&N^{2}G_{2c}(z,w) = - {1\over 4}{\partial \over{\partial
w}}{\partial
\over{\partial z}} {\rm Log} [ 1 - G^{2}(z) G^{2}(w) ]{\nonumber}\\
   &=& ({1\over{G(z)G(w)}} - G(z)G(w))^{-2}{1\over{G(z)G(w)}}
({\partial G(z) \over{\partial z}})({\partial G(w)\over{\partial w}})
\end{eqnarray}
This leads to
%(2.19)
\begin{equation}
 N^{2}G_{2c}(z,w) = {1\over{4(z^{2} - w^{2})^{2}}}
[{2 z^{2}w^{2} - 4 z^{2} - 4 w^{2}
\over{\sqrt{z^{2} - 4} \sqrt{w^{2} - 4}}} - 2 z w ]
\end{equation}
%************************************
%In the limit $z \rightarrow w$, it gives
%{2.20}
%\begin{equation}
%  N^{2} G_{2c}(z,z) = {1\over{16 z^{2}(z^{2} - 4 )^{2}}}
%\end{equation}
%Note the divergence at $z=0$. This should be
%compared to the
%Hermitian case, in which $G_{2c}(z,z)$ remains finite except at
%the endpoints of the spectrum
%$z$ = 2 and $-2$. We also note that (2.18) does not coincide with
%the
%universal correlation function for a  Hermitian matrix model, but
%it agrees with the generalized universal form given in [4].

 The two-point correlation function $\rho_{2c}(z,w)$
agrees with the universal behavior in the short distance $z
\rightarrow
w$ limit:
%(2.21)
\begin{equation}
   \rho_{2c}(z,w) = - {1\over{2 \pi^{2} N^{2} (z - w)^{2} }}
\end{equation}

  By the equation of motion method, the complex matrix model
can be
studied
using a recursion approach order by order in $1/N^{2}$ [11].
Pole terms appear in each order of $1/N^{2}$. The  one-point
Green's
function diverges in the $k$-th order as
%{2.22}
\begin{equation}
    \delta G(z) \sim {c_{k}\over{N^{2k}z^{2k}}}
\end{equation}

Since  the connected two-point
correlation
function in the Hermitian matrix model does not have a double
pole in the
orthogonal polynomial approach, let us use the orthogonal
polynomial
approach to investigate whether we have the same
situation for the one-point Green's function $G(z)$ in the
limit $z \rightarrow 0$.

\vskip 5mm

%***********************************************
%  section 3
%***********************************************
\sect{ Orthogonal polynomial analysis }
%************************************************

   Since $\varphi^{+}\varphi$ can
be regarded as a Hermitian matrix with positive
eigenvalues, we can write the joint
probability distribution
in terms of its positive eigenvalues $\epsilon_{i}$ [19,20]
%{3.1}
\begin{equation}
   P_{N}(\varepsilon_{1},...,\varepsilon_{N})d \varepsilon_{1} \cdots
d \varepsilon_{N} = C e^{- N \sum \varepsilon_{i}}\Pi_{i<j}
(\varepsilon_{i} - \varepsilon_{j})^{2}d \varepsilon_{1}\cdots d
\varepsilon_{N}
\end{equation}
with $C$ a normalization constant.
The relevant orthogonal polynomial is the Laguerre polynomial
$L_{n}(x)$, where $L_{0}(x) = 1$, $L_{1}(x) = 1 - x$, and $L_{2}(x) =
1 - 2 x + x^{2}/2$.
The density of state $\rho(\varepsilon)$
and the two-point connected correlation function\\
$\rho_{2c}(\varepsilon,\varepsilon')$
are given in terms of the kernel $K(\varepsilon,\varepsilon')$
 [1] as
%(3.2)
\begin{equation}
   \rho(\varepsilon) = K(\varepsilon,\varepsilon)
\end{equation}
%(3.3)
\begin{equation}
   \rho_{2c}(\varepsilon,\varepsilon') = - [K(\varepsilon,\varepsilon')]^{2}
\end{equation}
where
%(3.4)
\begin{equation}
     K(\varepsilon,\varepsilon') =
{1\over {N}} \sum_{0}^{N-1}\psi_{n}(\varepsilon)\psi_{n}(\varepsilon')
\end{equation}
and
%(3.5)
\begin{equation}
    \psi_{n}(\varepsilon) = e^{-N\varepsilon/2}L_{n}(N\varepsilon)
\end{equation}
Noting that $\varepsilon = E^{2}$ in our terminology, we have an
extra
factor of $E$ for the density of state since $d\varepsilon = 2 E dE$. At
$E=0$, all
Laguerre polynomials are equal to one, so the density of state
should vanish
at $E = 0$ due to this extra factor $E$ in accordance with the
reflection
symmetry argument given earlier.

This result for the density of state at $E = 0$ seems to contradict
the result we obtained in the previous section. But there is no
contradiction.
In the previous section, we took the large $N$ limit first, so that the
density is effectively smoothed over a certain width.
Here we take
$N$ large but fixed and let  $E$ go to zero, and obtain $\rho(E) =
0$. We
see the hole described earlier. This non-commutativity of the two
limits
discussed here is  similar to the discussion of the two-point
connected
correlation
function of a Hermitian matrix model [1, 2].

Using the Christoffel-Darboux identity, we have a compact
expression
for the density of state,
%(3.6)
\begin{eqnarray}
  \rho(\varepsilon) &=& e^{-N\varepsilon} \sum_{k=0}^{N-1}
L_{k}^{2}(N\varepsilon)\cr
       &=& N e^{-N\varepsilon} [ L_{N}(N\varepsilon) L_{N-1}^{'}(N\varepsilon)
 - L_{N-1}(N\varepsilon) L_{N}^{'}(N\varepsilon) ].
\end{eqnarray}
Since $ \varepsilon = E^{2}$, the density of state $\rho(E)$ is
\begin{equation}
\rho(E) = 2 E e^{- NE^{2}} \sum_{k=0}^{N-1} L_{k}^{2}(NE^{2})
\end{equation}
In Fig. 2, this density of state is shown for $N = 5$ and $N = 10$.
There appears N-oscillations in the density of state, and the first peak
near  $E = 0$ is finite for $N \rightarrow 0$.
The ratio of the value of the first peak to the second one is almost 1.2.
In Fig.2, the dotted line represents the semi-circle behavior in the
large N limit given by $\sqrt{ 4 - E^2} / \pi$. When the oscillating part
is averaged smoothly, we obtain the correction of the density of state
to the semi-circle law of order $1/N^2$.
\begin{equation}
    \Delta \rho = < \rho(E) > - {1\over \pi} \sqrt{ 4 - E^2 }
\end{equation}

It may be useful to write the asymptotic expression by the Bessel function.
 Knowing the large $N$ behavior (for $E$  small ) of Laguerre
polynomials, we can
write the preceding in terms of
 Bessel function. Remarkably, the oscillating part near $E = 0$ in Fig.2
is approximated by
%(3.7)
\begin{equation}
  \rho(E) \simeq  2NE [J_{0}^{2}(2NE) + J_{1}^{2}(2NE)]
\end{equation}
By plotting the oscillating curve of this equation, we find that the
first peak near $E = 0$ is almost same as the first peak value of (3.7).
The ratio of the first peak to the second one is also 1.2, which we mentioned
before.
 Now we use Hankel's asymptotic expansion of
 Bessel functions $J_{0}(t)$, $J_{1}(t)$, for the large $t = 2NE$.
%(3.10)
\begin{eqnarray}
J_{0}(t) &=& \sqrt{{2\over{\pi t}}}[P(0,t){\rm cos}
(t - {\pi\over{4}})
 -
Q(0,t){\rm sin}(t - {\pi\over{4}})],\nonumber\\
J_{1}(t) &=& \sqrt{2\over{\pi t}} [ P(1,t){\rm cos}(t - {3\pi\over{4}})
- Q(1,t){\rm sin}(t - {3\pi\over{4}})]
\end{eqnarray}
where
%(3.11)
\begin{eqnarray}
   P(l,t) &=& \sum_{k=0}^{\infty} (-
1)^{k}{(l,2k)\over{(2t)^{2k}}}\nonumber\\
          &\sim& 1 - {(4l^2 - 1) (4l^2 - 9)\over{128 t^2}} + \cdots\nonumber\\
 Q(l,t) &=& \sum_{k=0}^{\infty} (-1)^{k} {(l,2k+1)\over{
(2t)^{2k+1}}}\nonumber\\
          &\sim& {4l^2-1 \over{8t}} + \cdots\nonumber\\
 (l,m) &=& {\Gamma({1\over{2}} + l + m)\over{m!
\Gamma({1\over{2}}+ l -m)}}
\end{eqnarray}

We use the smooth average for the oscillating part by setting $< {\rm sin}^2
(x) > \\
= < {\rm cos}^2 (x) > = {1\over {2}}$. Then it is easy to find that
the density of state becomes
\begin{equation}
\rho^{\rm smooth}(E) = {2\over{\pi}}( 1 + {1\over{32N^{2}E^{2}}} -
{9\over{2048N^{4}E^{4}}}
+ \cdots ).
\end{equation}

The term of order $1/N^{2}$ agrees with the result obtained in the
previous section by the
diagrammatic method. In this approximation, the leading term is given
simply by ${2\over{\pi}}$ for ${\sqrt{4 - E^2}/{\pi}}$.
Thus we find that the singular double pole at order $1/N^{2}$ is in
a sense
spurious, which appears because we took the large $N$ limit first,
and it is recovered after the smooth average [1]. The density of state
does not diverge for $E \rightarrow 0$.
Recently the density of state of this complex matrix model has
been
studied in various contexts, and related expressions in terms of
Bessel
functions
have been discussed [22].

We note that in the case of the Hermitian matrix model, the
Bessel
function is of
half-integer order, and consequently we have
no poles in the $1/N$ expansion.
The double pole $1/x^{2}$ of the
connected two-point correlation function cancels with a factor
${\rm
sin^{2}}
(x)$ as $x \rightarrow 0$ [1].
Here we have a different situation since ${\rm sin^{2}}
(x - {\pi\over{4}})$ does not vanish for $x \rightarrow 0$. We must
sum the leading pole terms of the equation above and it gives
eventually a finite result for $E \rightarrow 0$ at fixed large $N$.

%**************************************************
%  section four
%**************************************************
\sect {Lattice of matrices}
  We now extend our analysis to the
general $d$-dimensional lattice of matrices [3]. We place $N$ by
$N$
matrices
on the lattice. The gluon propagator is given by $\sigma_{\alpha
\beta}$
defined by
%(4.1)
\begin{equation}
    \sigma_{\alpha \beta} = {1\over{M_{\alpha \beta}^{2}}}
\end{equation}
Here $M_{\alpha \beta}$, as defined in [3], is a real symmetric
matrix
whose entries are the analogs of $m_1^2$, $m_2^2$, and
$m^2$ in (2.6).
The
indices $\alpha$ and $\beta$, which label the lattice sites, run
from $1$ to
$C$, where $C$ denotes the number of sites on the lattice. In
leading
order, the one-point Green's function is given by $G(z) = {1\over N}
\sum_{\alpha} g_{\alpha}$ where $g_{\alpha}$ is determined by
%(4.2)
\begin{equation}
   g_{\alpha} = {1\over{ E - \sum_{\beta} \sigma_{\alpha \beta}
g_{\beta}}}
\end{equation}

Let us restrict ourselves to a
$d$-dimensional hypercubic lattice, on which
a quantum particle hops with a hopping matrix is given by
$\sigma$.
We introduce $\varepsilon(k)$ by [3]
%(4.3)
\begin{equation}
   \sigma_{\alpha \gamma} = \sum_{k}< \alpha\vert k >
\epsilon (k) < k  \vert \gamma >
\end{equation}
For the case of nearest neighbor hopping, we have
%(4.4)
\begin{equation}
    \varepsilon (k) =
{1\over{m^{2}}} + {2\over{M^{2}}} \sum_{a} {\rm cos}
     k_{a}
\end{equation}
Below we will calculate the one-point Green's function for the most
general
case with arbitrary $ \varepsilon (k)$. For specific examples, we
often take for simplicity the case $m^2 = \infty$ and $M^2 = 2$.

  We will now calculate the one-point Green's function to order
$1/N^{2}$. We consider
here the general situation defined by some $\sigma$ matrix. In
particular,
for the simple example given in (2.5) we include the diagonal
part of the Hamiltonian.
The self-energy part to order $1/N^{2}$ is obtained from the
diagram of $D_{a}$ and $D_{b}$, where the number of rungs on
the
ladders
are
no longer restricted as they were before. We obtain
%(4.5)
\begin{equation}
   \Sigma_{a} = G [({\sigma G^{2}\over{1 - \sigma G^{2}}})_{\alpha
\alpha}]^{2}
 + G [{\sigma^{2}G^{4}\over{(1 - \sigma G^{2})^{2}}}]_{\alpha
\alpha}
({\sigma G^{2}\over{1 - \sigma G^{2}}})_{\beta \beta}
\end{equation}
Note that repeated indices are not summed unless indicated
otherwise. By translation invariance this expression is actually
independent of $\alpha$ and $\beta$.

 For the self-energy $\Sigma_{b}$, we have two parts
$\Sigma_{b1}$
and $\Sigma_{b2}$. We obtain for $\Sigma_{b1}$
%(4.9)
\begin{equation}
  G \sum_{\gamma} (\sigma^{n})_{\gamma \alpha}
(\sigma^{m})_{
\alpha \gamma} (\sigma^{k})_{\gamma \alpha} G^{2 n} G^{2 m}
G^{2 k} = G \sum_{\gamma} [ ({\sigma G^{2}\over{1 - \sigma
G^{2}}})_{
\alpha \gamma}]^{3}
\end{equation}
where $n$, $m$ and $k$ are the number of the gluon
propagators.
The other part $\Sigma_{b2}$,  obtained by cutting one
quark
propagator inside the ladder, becomes
%(4.10)
\begin{eqnarray}
 &&G \sum_{\beta \gamma} \sum_{n,m,j,k}^{\infty}
  (\sigma^{m})_{\beta \gamma} (\sigma^{n})_{\gamma \alpha}
   (\sigma^{k})_{\gamma \beta} (\sigma^{j})_{\beta \alpha}
    G^{2n} G^{2m} G^{2j} G^{2k}\nonumber\\
 &=& G \sum_{\beta \gamma} [ ( {\sigma G^{2}\over{1 - \sigma
 G^{2}}} )_{\beta \gamma}]^{2} ({\sigma G^{2}
  \over{1 - \sigma G^{2}}} )_{\alpha \gamma}
   ({\sigma G^{2}\over{1 - \sigma G^{2}}} )_{\alpha
\beta}
\end{eqnarray}
Note the rather unusual wasy in which the indices or site labels
are arranged.

Let us check these expression for the case $C = 2$ discussed
earlier. We set
%(4.6)
\begin{equation}
     \sigma = \left (\matrix{ 0 & 1\cr
                        1& 0}\right )
\end{equation}
The first term in (4.5) becomes
%(4.7)
\begin{equation}
      ( {\sigma G^{2}\over{1 - \sigma G^{2}}} )_{\alpha
\alpha } = {1\over{2}}( {G^{2}\over{ 1 - G^{2}}} - {G^{2}
\over { 1 + G^{2}}} ) = {G^{4}\over{ 1 - G^{4}}}
\end{equation}
and
%(4.8)
\begin{equation}
    (  ( {\sigma G^{2}\over{1 - \sigma G^{2}}} )^{2}
 )_{\alpha \alpha} = {1\over {2}} ( {G^{4}\over{( 1 - G^{2})
^{2}}}+ {G^{4}\over{( 1+ G^{2})^{2}}} ) = {G^{4}( 1+ G^{4})\over
{( 1 - G^{4} )^{2}}}
\end{equation}
 The factor $1/2$ in (4.7) and (4.8) are necessary for a fixed
$\alpha$ (which we do not sum over.)
Thus we get the same result for the self-energy
$\Sigma_{a}
= 2 G^{9}/(1 - G^{4})^{3}$ as was given earlier
(2.11).
As for $\Sigma_b$, we note
%(4.11)
\begin{eqnarray}
   &&(\sigma^{n})_{\alpha \gamma} = \delta_{\alpha \gamma}
{\rm (n
= even)}
  \nonumber\\
  && (\sigma^{n})_{\alpha \gamma} = \sigma_{\alpha \gamma}
{\rm
(n=odd)}
\end{eqnarray}
Using this property, we see that the numbers of gluon
propagators
$n,m,m$ or $n+j,m,k$ should be either all even or all odd. Thus
we
obtain the previous result  for $\Sigma_{b}$ as given in (4.9) and
(4.10).

\vskip 5mm
We can now immediately go to a $d$-dimensional lattice on
which matrices
are
placed by inserting the $\sigma$ given in (4.3). Thus, we obtain
%(4.12)
\begin{eqnarray}
   ({\sigma G^{2}\over {1 - \sigma G^{2}}})_{\alpha \beta}
   &=& <\alpha \vert {\sigma G^{2}\over{1 - \sigma G^{2}}}\vert
\beta >\nonumber\\
   &=& \sum_{k} <\alpha \vert k><k \vert {\sigma G^{2}\over{
  1 - \sigma G^{2}}}\vert k><k\vert \beta>\nonumber\\
   &=& {1\over C}\sum_{k} e^{i
k(\alpha - \beta)} ({\varepsilon_{k}
G^{2}\over{1 - \varepsilon_{k} G^{2}}})
\end{eqnarray}
where we have used $<\alpha \vert k><k\vert \beta> = {1\over
C}e^{i
k(\alpha - \beta)}$. Using this diagonalized representation, we
obtain for the different parts of the self-energy the following
expressions:
%(4.13)
\begin{equation}
\Sigma_{a1} = G [ {1\over {C}}\sum_{k} {\varepsilon_{k} G^{2}\over
{
1 - \varepsilon_{k}G^{2}}}]^{2}
\end{equation}
%(4.14)
\begin{equation}
    \Sigma_{a2} = G ({1\over{C}}\sum_{k}
{\varepsilon_{k}^{2}G^{4}\over
{ (1 -
\varepsilon_{k}G^{2})^{2}}})({1\over{C}}\sum_{p}{\varepsilon_{p}G
^
{2}
\over{1 - \varepsilon_{p}G^{2}}})
\end{equation}
%(4.15)
\begin{equation}
  \Sigma_{b1} = G \sum_{\gamma}[{1\over{C}}
\sum_{k}e^{ik(\alpha -
\gamma)}
({\varepsilon_{k}G^{2}\over{1 - \varepsilon_{k}G^{2}}})]^{3}
\end{equation}
%(4.16)
\begin{eqnarray}
  \Sigma_{b2} &=& G \sum_{\beta,\gamma}[ ({1\over{C}}\sum_{k}{
\varepsilon_{k}G^{2}\over{1 - \varepsilon_{k}G^{2}}}e^{ik(\beta -
\gamma)})^{2}
({1\over{C}}\sum_{q}{\varepsilon_{q}G^{2}\over{1 -
\varepsilon_{q}G^{2}}}
e^{iq(\alpha - \gamma)})\nonumber\\
&\times&({1\over{C}}\sum_{p}{\varepsilon_{p}G^{2}\over{1 -
\varepsilon_{p}G^{2}}}e^{ip(\alpha - \beta)})]
\end{eqnarray}
where $C$ is the number of lattice points, $C=L_{1}\cdots L_{d}$.
We note that by translation invariance $\Sigma_{b1, b2}$ are in
fact independent of $\alpha$.

\vskip 5mm
We can easily recover our previous results of course. For the two-site case
we have $ k=0$ and $k=\pi$, and $\beta, \gamma,
= 1,2$ with a fixed $\alpha=1$.  For instance, noting that
$\varepsilon_{k=0}=1$, and $
\varepsilon_{k=\pi}= -1$, we get immediately
%(4.17)
\begin{eqnarray}
  \Sigma_{b1} &=& {G\over{8}}
\sum_{\gamma=1}^{2} ( {G^{2}\over{ 1 - G^{2}}}
  - e^{i\pi(1 - \gamma)}{G^{2}\over{1 + G^{2}}})^{3}\nonumber\\
  &=& {G^{7}( 1 + G^{6})\over{( 1 - G^{4})^{3}}}
\end{eqnarray}
in agreement with(2.12). Similarly, we find easily that
$\Sigma_{b2}$
calculated here agrees with (2.13).

 We have no divergence for the 3-site lattice, where  $k = 0,
{2\pi\over
{3}},{4\pi\over{3}}$ and $\varepsilon_{k}= 1, - {1\over{2}}, -
{1\over{2}}$,
respectively. Similarly for lattices with odd number of sites.

In contrast, for a one-dimensional lattice with the number of sites
$C=L=$
an even
integer, we have $k =
\pi$ and
$\varepsilon_{k=\pi} = -1$, and thus we get a divergence when
$G^{2}=-
1$.

For $L\rightarrow \infty$,
we find
%(4.19)
\begin{eqnarray}
   \Sigma_{a1} &=& G [ {1\over{C}}\sum_{k} {\cos (k) G^{2}\over { 1
-
\cos (k)
G^{2}}}]^{2}\nonumber\\
   &=& G{(1 - \sqrt{1 - G^{4}})^{2}\over{1 - G^{4}}}
\end{eqnarray}
%(4.20)
\begin{equation}
   \Sigma_{a2} = G {(2G^{4} - 1 + ( 1 - G^{4})^{3/2})(1 - \sqrt{1 -
G^{4}})
\over{ ( 1 - G^{4})^{2}}}
\end{equation}
where $k=2\pi (j-1)/L, j=1,\cdots L$
For the self-energy parts $\Sigma_{b1},\Sigma_{b2}$, the sum
over
$\beta$ and $\gamma$ give complications. But the leading
singularity of
$\Sigma_{b2}$ cancels exactly the singularity of $\Sigma_{a2}$ in
(4.20). The singularity of $\Sigma_{b1}$ at $G^{2}=-1$ is the same
as the
singularity of $\Sigma_{a1}$.
 Thus we determine the singularity for one-dimensional case  at
the
band center $G^{2}=-1$ with
a single pole divergence in order $1/N^{2}$.
The cancellation of the leading singularity of $\Sigma_{b2}$ with
$\Sigma_{a2}$
holds for any dimension and coincides with the cancellation
found in ref.[5]
for the lowest Landau level. We have also verified this
cancellation
by
the numerical evaluation of the self-energy
for large $L$ near $G^{2}= -1$.

  The singularity $1/ N^{2}\vert E \vert$
in the density of state as $E \rightarrow 0$ is too strong,
since the integral of the density of state
should be equal to one. In one dimension, the off-diagonal
disorder
case of the tight binding model, in which the hopping matrix is real and
$N = 1$ case,
 is known to have a singularity of
$1/\vert E \vert ( \ln \vert E \vert )^{3}$
in the density of state
near  $E=0$ [14]. In Fig.3, the density of state of a finite chain with
real hopping matrix for N=1, is evaluated for a box distribution.
We consider the nearest neighbour random hopping, which is represented by
the tridiagonal real matrix $M$,
\begin{equation}
   M = \left(\matrix{0&b_{1}^{*}& 0 & \dots\cr
          b_{1}& 0& b_{2}^{*}& 0& \dots\cr
          0& b_{2}& 0 & b_{3}^{*} & \dots \cr
          0 & 0& b_{3} & 0 & b_{4}^{*}  \cr
           &  & \dots \cr}
\right).
\end{equation}
By the calculation of the density of the eigenvalue of this matrix,
we obtain the curve of  the density of state. The random variable $b_{i}$
is generated 5000 times, and the histogram of the eigenvalues is evaluated.
{}From these calculation, we see the divergent singularity near E = 0.
The singularity is consistent with $1/E$ behavior, although it is difficult
to see the existence of the logarithmic correction from this calculation.

For the complex hopping case, in which the coupling $b_{i}$ in (4.20)
 is a complex random number, the density of state shows oscillations similar
to Fig.2. We evaluate this complex case in Fig. 4. The first peak near
$E = 0$ is finite for $L \rightarrow \infty$, where $L$ is the length
of the chain.

It may be also interesting to note that we have the similar behavior in
different examples. For the sparse random matrix, the density
of state shows a singularity  $1/\vert E \vert  (\ln \vert E \vert )^{2}$
instead of $1/\vert E \vert (\ln
\vert E \vert )^{3}$
 [15].
We have also another example, studied by Br\'ezin, Gross and
Itzykson who have obtained the same singularity
$1/\vert E \vert (\ln (\vert E \vert ))^{2}$ in the density of state for
the lowest Landau level with a random Poisson distribution [16].

For the two dimensional case, the self-energy $\Sigma_{a1}$
behaves like $[\ln (
1 +
G^{2})]^{2}$ and $\Sigma_{a2}$ like $\ln (1 +G^{2})/(1 + G^{2})$.
This singularity is cancelled with a singularity in $\Sigma_{b2}$.
We find that
the singularity of the self-energy  is proportional to $[\ln (1 +
G^{2})]^{2}$
near the band center. This coincides with the calculation of the
two-site lowest Landau level result in ref. [5].

It is remarkable that we have the same singularity as in the two-
state
(spin up and down) degenerate quantum Hall case studied in [5].
We consider this a manifestation of the possibility that these two
mdoels may same universality class.
The integral of the density of state with this logarithmic square
singularity is finite. The exact singularity of
the off-diagonal tight binding model is not known.
Numerical work [9] for the two-state lowest Landau level model
also shows
a singularity at the band center also. Moreover,
the logarithmic square singularity
seems to have an applicable region near $E = 0$ according to
a recent
numerical study for the lowest Landau level, although the density of
state does not diverge for $E \rightarrow 0$ [23].
 The state of $E=0$ in the two-dimensional case is related to the
zero energy
wave function, for which the Atiyah-Singer index theorem can be
applied.
In this respect, one can perhaps relate the present problem to
other
interesting problems [24,25].

  For two-dimensional case, we have also evaluated the density of state
of the off-diagonal disorder (N = 1). We examined the real and
complex case similar to the one dimensional case. For the real case,
the density of state seems divergent at $E = 0$ and consistent with
[26]. However, the density of state of the  complex case,
in which nearest neighbour hopping
matrix element is complex, shows the similar behavior to the complex
matrix model (Fig.2). The first peak near $E=0$ is finite.

%**************************************************
%  Discussion
%*************************************************
\sect{Discussion}
%**************************************

 In this paper, we have discussed the singularity of the density of
state
in the large $N$ limit of a complex matrix model as well as
its $d$-dimensional lattice generalization.
We have found the singularity in order $1/N^{2}$.
We have compared our result with known results in $d=0$ and
$d=1$.
For  $d=0$, the double pole singularity we obtained is reminiscent
to
the spurious double pole of the two-point correlation function in
the
Hermitian matrix model.
For $d=1$, we compared our result with the off-diagonal tight
binding model. Our result is different by the logarithmic factor.
For $d=2$,  we have obtained the logarithmic square
singularity, which coincides with the result of [5] for the
 $N$-orbital two-state lowest Landau level quantum Hall model.
We interprete this coincidence of the singularity as a
manifestation of the possibility that these models belong to the
same universality case. Although our
calculation
is restricted to order $1/N^2$, the result of the singularity of the
density
of state gives a clue of what the true behavior might be. It may be
interesting to evaluate the Green's functions to order
$1/N^4$.
 For the spin degenerate two-state quantum Hall system, the numerical
similation suggests that there are  three extended states [9]. In $1/N$
expansion, the singularity $( {\rm ln} E )^{2}/N^{2}$
has been obtained for the density of state [5]. However, as we discussed
in this paper, this is interpreted as the result after the
smooth average. The density of state is considered to be finite
for $E \rightarrow 0$. It is also important to note that the density of state
has a nonvanishing value for $E \rightarrow 0$. Then, as discussed in [5],
the conductivity is exactly given by $\sigma_{xx} = e^2/\pi h$ since
the parameter $\theta$ in [5], defined by $\theta = - {\rm tan}^{-1}({\rm
Im}G(z)/
{\rm Re}G(z))$, becomes $- \pi/2$.

%**********************************************
\vskip 5mm
\begin{center}
{\bf Acknowledgement}
\end{center}
We thank Edouard Br\'ezin for discussions about the formulation of
lattices of matrices. SH thanks K. Minakuchi for discussion of his
numerical results. AZ would like to thank D.
Arovas for mentioning that the lattices of
matrices studied in [3] may be relevant for impurity scattering in
spin-dependent quantum Hall systems. He would also like to
thank M. Kohmoto
for hispitality at the Institute for Solid State Physics, University of
Tokyo,
where this work was initiated. This work is supported in part by the
National Science Foundation under Grant No. PHY89-04035, and by CNRS-JSPS
cooperative research project.

%************************************************************
%   references
%************************************************************
\newpage
\begin{center}
{\bf References }
\end{center}
\vskip 3mm

\begin{description}

\item[{[1]}] E. Br\'ezin and A. Zee, Nucl. Phys. {\bf B40} (1993), 613.
\item[{[2]}] E. Br\'ezin and A. Zee, Phys. Rev. {\bf E49} (1994) 2588.
\item[{[3]}] E. Br\'ezin and A. Zee, "Lattice of matrices," Santa
Barbara preprint
             (1994) NSF-ITP-94-75.
\item[{[4]}] E. Br\'ezin, S. Hikami and A. Zee,
             Phys. Rev. E (1995), in press. (hep-th.9412230).
\item[{[5]}] S. Hikami, M. Shirai and F. Wegner, Nucl. Phys. {\bf
B408} (1993) 415.
\item[{[6]}] R. M. Gade, Nucl. Phys. {\bf B398} (1993), 499.
\item[{[7]}] D. K. K. Lee, Phys. Rev. {\bf B50} (1994), 7743.
\item[{[8]}] A. W. W. Ludwig, M. P. A. Fisher, R. Shanker and G.
Grinstein,
             Phys. Rev. {\bf B50} (1994), 7526.
\item[{[9]}] C. B. Hanna, D. P. Arovas, K. Mullen and S. M. Girvin,
             a preprint (1994). cond-mat.9412102.
\item[{[10]}] F. J. Wegner, Phys. Rev. {\bf B19} (1979), 783.
\item[{[11]}] Yu. M. Makeenko, JETP-Lett. {\bf 52} (1990), 259.\\
              J. Ambj\o rn, C. F. Kristjansen and Yu. M. Makeenko,
             Mod. Phys. Lett. {\bf A7} (1992), 3187.
\item[{[12]}] R. Oppermann and F. Wegner, Z. Phys. {\bf B34}
(1979), 327.
\item[{[13]}] S. Hikami, Prog. Theor. Phys. {\bf 72} (1984), 722;
               Prog. Theor. Phys. {\bf 76} (1986), 1210.
\item[{[14]}] G. Theodorou and M. H. Cohen, Phys. Rev. {\bf B13}
          (1976), 4597.\\
  T. P. Eggarter and R. Riedinger, Phys. Rev.
          {\bf B18} (1978), 569.
\item[{[15]}] J. Rodgers and C. De Dominicis, J. Phys. {\bf A23}
              (1990) 1567.
\item[{[16]}] E. Br\'ezin, D. J. Gross and C. Itzykson, Nucl. Phys. {\bf
B235}
     (1984), 24.
\item[{[17]}] X. G. Wen and A. Zee, Phys. Rev. Lett. {\bf 69} (1992) 953,
             3600(E), Phys. Rev. {\bf B47} ( 1993) 2265.\\
             X. G. Wen and A. Zee, " A phenomenological study of interlayer
             tunneling in double-layered quantum Hall systems", MIT-SBITP
             preprint.
\item[{[18]}] K. Yang, K. Moon, L. Zheng, A. H. MacDonard, S. M. Girvin,
             D. Yoshioka and
             S-C. Zhang, Phys. Rev. Lett. {\bf 72} (1994) 732.
\item[{[19]}] K. Slevin and T. Nagao, Phys. Rev. Lett. {\bf 70} (1993) 635.
\item[{[20]}] B. V. Bronk, J. Math. Phys. {\bf 6} (1965), 228.
\item[{[21]}] M. Abramowitz and I. A. Stegun,
      Handbook of Mathematical Functions, National Bureau of
Standards,
      (1972) Washington, P.364.
\item[{[22]}] T. Nagao and K. Slevin, J. Math. Phys.
       {\bf 34} (1993) 2075.\\
         A. V. Andreev, B. D. Simons and N. Taniguchi,
              Nucl. Phys. {\bf B432[FS]} (1994) 487.
\item[{[23]}] K. Minakuchi, Master thesis, University of Tokyo (1995).
\item[{[24]}] X. G. Wen and A. Zee, Nucl. Phys. {\bf B316} (1989), 641.
\item[{[25]}] M. Inui, S. A. Trugman and E. Abrahams, Phys. Rev. {\bf B 49}
(1994) 3190.
\item[{[26]}] S. N. Evangelou, J. Phys. {\bf C 19} (1986), 4291.
\end{description}

%******************************************************
\newpage
%****************************************************
% Figure caption
%**********************************************
\begin{center}
{\bf Figure caption}
\end{center}
\vspace{5mm}
\begin{description}
\item[Fig.1a]
The diagrams $D_{a}$ and $D_{b}$ of order $1/N^2$.
\item[Fig.1b]
The self-energy $\Sigma_{a1},\Sigma_{a2},\Sigma_{b1}$ and $\Sigma_{b2}$,
which are obtained from $D_{a}$ and $D_{b}$ by cutting the solid lines.
\item[Fig.2]
The density of state of the complex matrix model for $N=5$ (broken line)
 and $N=10$ (solid line). The semi-circle law $\sqrt{4 - E^2}/\pi$ is
also shown by a dotted line.
\item[Fig.3]
The density of state of a finite chain model (L = 10)
 with off-diagonal disorder.
The hopping random variable $b_{i}$ is real and obeys the box distribution
with $-1 < b_{i} < 1$.
There is a divergence at $E = 0$.
\item[Fig.4]
The density of stae of a finite chain model (L = 12)
 with off-diagonal disorder.
The hopping random variable $b_{i}$ is complex.
The real part and  the imaginary part
both obey the box distribution, which takes the value between $-1$ and 1
uniformly.
\end{description}
\end{document}